\documentclass[aps,prd,preprint,groupedaddress]{revtex4}

\usepackage{epsfig}
\usepackage{latexsym}
\usepackage{amsmath}
\include{mssymb}
\include{macros}

\begin{document}

\def\beq{\begin{equation}}
\def\eeq{\end{equation}}
\def\bea{\begin{eqnarray}}
\def\eea{\end{eqnarray}}

\preprint{DESY 04-142}
\title{Low $\boldmath x\unboldmath$ particle spectra in the Modified Leading Logarithm Approximation}
\author{S. Albino}
\affiliation{{II.} Institut f\"ur Theoretische Physik, Universit\"at Hamburg,
             Luruper Chaussee 149, 22761 Hamburg, Germany}
\author{B. A. Kniehl}
\affiliation{{II.} Institut f\"ur Theoretische Physik, Universit\"at Hamburg,
             Luruper Chaussee 149, 22761 Hamburg, Germany}
\author{G. Kramer}
\affiliation{{II.} Institut f\"ur Theoretische Physik, Universit\"at Hamburg,
             Luruper Chaussee 149, 22761 Hamburg, Germany}
\date{\today}
\begin{abstract}
We show that the higher moments of the evolution obtained
from the Modified Leading Logarithm Approximation may be regarded as
spurious higher order terms in perturbation theory, and that neglecting them 
leads to a good description of the data around and above the peak in 
$\xi=\ln (1/x)$. Furthermore, we use this study of the moments to show that 
at high energy the Limiting Spectrum with Local Parton-Hadron Duality may also be derived 
from the Modified Leading Logarithm Approximation without
any non-perturbative assumptions.
\end{abstract}
\maketitle


\section{Introduction}
\label{Intro}

The perturbative QCD approximation is consistent with a wide range of
data. However, there are two formal limitations on the predictive power of
this approximation. Firstly, perturbative 
QCD is incomplete in that it does not describe the physics of hadrons entirely. Secondly, 
the perturbation series becomes singular when 
any of the virtual and real quarks and gluons (collectively called partons) in a process 
has a configuration of energy and momenta whose ``energy scale'' $E$, a 
quantity whose definition cannot be precisely
defined and which is integrated over in virtual loops,
is as small as $\Lambda_{\rm QCD}$, the fundamental scale of QCD which determines
the scale at which non-perturbative effects become important.
For virtual partons, these two problems are related through the
factorization theorem, which states that the dominant (leading twist) contribution to
a hadronic process at high energy is given by a convolution
over dimensionless kinematic variables of process dependent quantities describing those partons
with $E$ greater than the factorization scale $Q$, which is fixed, with
process independent quantities which describe both asymptotic hadrons and partons with $E<Q$. 
The former quantities and the $Q$ dependences of the latter quantities
are calculable in perturbation theory in a given factorization scheme,
which defines $E$, provided that
$Q$ is sufficiently much larger than $\Lambda_{\rm QCD}$. Processes with real
partons are calculable in perturbative QCD provided one performs a
physical sum over these particles.

In the fixed order approach, where the perturbation series is calculated to 
a finite order, the perturbative series for a given partonic process can become
divergent in certain regions of phase space. Fortunately, the
terms that cause the divergences in a given region are often calculable to all orders, with
the formal sum of all terms of a given class being free of divergences in that region and forming
the term of a given order in a new series.

The hadronic and low $E$ partonic components of processes involving
the inclusive production of hadrons are contained in fragmentation
functions (FF's), which describe the probability of transition from a
high $E$ parton $a$ to a low $E$ hadron $h$. 
In this case it is valid to identify $E$
with the transverse momentum of each parton. The
cross section for a given process is obtained by convoluting the FF's
with the partonic cross sections (the coefficient functions) over the 
ratio $z$ of the hadron's longitudinal momentum to that of the parton.
The FF's at one value of $Q$ can be calculated perturbatively 
from those at another value $Q_0$ by convoluting with the evolution matrix, which describes the
probabilities of transition from a parton at $Q$ to another at $Q_0$, and which is 
obtained in terms of the perturbatively calculable splitting
functions by solving the DGLAP equation. However, the series for the
splitting functions breaks down as $z\rightarrow 0$ due to terms which
behave in this limit like $(\alpha_s^{n}/z) \ln^{2n-1-m}z$, where
$m=1,...,2n-1$ labels the class of terms. (Terms which behave like $\alpha_s^{n}$
are classified as $m=2n$.) These logarithms must be resummed 
before the fixed order splitting functions are valid at
small $z$. Since the cross section at hadronic momentum fraction $x=2p/\sqrt{s}$,
where $p$ is the momentum
of the produced hadron $h$ and $\sqrt{s}$ is the centre-of-mass energy, 
depends on the FF's over the range $x\leq z\leq 1$, 
such a resummation is required to describe the cross section at small $x$.
Resummation of the leading ($m=1$)
and subleading ($m=2$) logarithms, which appear at leading order, is obtained via the
Modified Leading Logarithmic Approximation (MLLA) \cite{Dokshitzer:1984dx} (for reviews see
\cite{Dokshitzer:1991wu,Khoze:1996dn}). Since the coefficient functions are non-singular as
$z\rightarrow 0$, and the quark FF's are proportional to the gluon FF
in the MLLA, the cross section at low $x$ is then proportional to the
MLLA evolved gluon FF. 

There is some freedom to choose the MLLA evolution
due to the next-to-MLLA error. The evolution given by the 
analytic solution to the MLLA differential equation \cite{Dokshitzer:1991ej} is well behaved for
$Q_0=O(\Lambda_{\rm QCD})$, which may imply that the two limitations on perturbation 
theory to describe hadronic physics that were mentioned at the 
beginning of this section are too stringent.
The first limitation may be weakened by introducing the 
Local Parton-Hadron Duality (LPHD) hypothesis \cite{Azimov:1984np}, which states
that the distribution of partons below a certain energy scale in
sufficiently inclusive processes is similar to the distribution of
hadrons, up to the number of particles actually produced (the
multiplicity). This implies that the initial hadronic fragmentation function 
is proportional to the partonic fragmentation function for $Q_0=O(\Lambda_{\rm QCD})$.
The second limitation may be weakened by assuming that
partons with $E=O(\Lambda_{\rm QCD})$ may be described by perturbation theory
(after resumming with the MLLA).
Together with the LPHD, the particular choice $Q_0=\Lambda_{\rm QCD}$ in the MLLA evolution 
gives the Limiting Spectrum \cite{Azimov:1984np}. The only free parameters are the initial normalisation and 
$\Lambda_{\rm QCD}$, which can be fitted to the data. A good fit to the data
over the whole range of $\xi=\ln (1/x)$ can be achieved, however only if
the evolution of the normalization is modified.
In \cite{Lupia:1997hj} an additional component not provided
by the MLLA was added to the normalization, whereas in
\cite{Abbiendi:2002mj} a different normalization was fitted for
each value of $\sqrt{s}$. Otherwise $\Lambda_{\rm QCD}$ obtained in this approach
is consistent with that of other analyses.

In our recent work \cite{Albino:2004xa}, we studied the MLLA evolution
without using strong assumptions such as the LPHD or the validity of
the Limiting Spectrum, nor modifying the MLLA evolution itself. Using
an initial scale $Q_0 \gg \Lambda_{\rm QCD}$ and a parameterized
function for the initial gluon FF, we achieved a good description of
the charged hadron cross section data for $\xi$ up to and around
the peak, and obtained values of $\Lambda_{\rm QCD}$ close to those in the
literature. Beyond the peak the MLLA evolution turned out not to be
sufficient to describe the data. The theoretical curves exhibited a
second bump after the first peak, not seen in the data, which have a
characteristic Gaussian shape around the first peak. Such a problem
cannot be solved by modifying the MLLA normalization. 

It is expected that
the fixed order approach with double and single logarithms resummed with the MLLA
should give a good description of the small to large $\xi$ data, and therefore,
if fixed order corrections are not included,
one has to consider qualitatively what effect the fixed order prediction
at small $\xi$ has on the MLLA prediction at large $\xi$. In fact,
the MLLA formally improves
the description of the Mellin transform of the cross section for small
$|\omega|$, where $\omega$ is defined in Eq.\ (\ref{melofD}), 
rather than the cross section itself at large $\xi$. 
Fixed order calculations indicate that the large $\xi$ region has rather little dependence on the
large $|\omega|$ region. 

It is the purpose of this paper to consider the features of the data
that the MLLA can describe, 
and thereby modify the MLLA evolution in order to improve the
description of the large $\xi$ behaviour of the spectra without
spoiling the description around the peak. We start in Section \ref{MitMLLA} by considering
the moments of the cross section, since these quantities depend very little on
the large $|\omega|$ behaviour of the evolution and
the parameterization of the initial gluon FF, and can be easily extracted from the data.
We then derive our approach for improving the large $\xi$ description.
In Section \ref{FtD}, we compare the predictions of this approach with the experimental data
at large $\xi$. In Section \ref{LPHDLS}, we study the effect of imposing
the limits of the LPHD and Limiting Spectrum on our approach. Finally, in Section
\ref{conclusions}, we present our conclusions.

\section{Evolution of moments in the MLLA}
\label{MitMLLA}

In this Section we outline the features of the MLLA that will be important for the
derivation of our main result. More details can be found in our previous publication \cite{Albino:2004xa},
however, to make our formulae here more transparent we will refrain from using
the variables $Y=\ln (Q/Q_0)$ and $\lambda=\ln (Q_0 /\Lambda_{\rm QCD})$
and write $Q$ and $Q_0$ explicitly.

The MLLA is believed to describe the energy dependence of cross sections for hadron production
in the region for which $\alpha_s \ll 1$ and $|\omega|=O(\sqrt{\alpha_s})$,
where $\omega$ replaces the variable $x$ in the Mellin transform
\beq
f_{\omega}=\int_0^\infty d\xi \exp[-\omega \xi]xf(x).
\label{melofD}
\eeq
In this 
limit the cross section is proportional to the gluon FF $D(x,Q)$ at the conventional choice 
of factorization scale $Q =\sqrt{s}/2$.
The dependence of the gluon FF on $Q$ in Mellin space takes the simple form
\beq
D_{\omega}(Q)=E_{\omega}(\alpha_s(Q),\alpha_s(Q_0))D_{\omega}(Q_0),
\label{defofanomdim}
\eeq
where $D_{\omega}(Q_0)$ is non-perturbative, while $E_{\omega}$ is determined in terms of
the gluon splitting function $\gamma_{\omega}(\alpha_s)$,
\beq
E_{\omega}(\alpha_s(Q),\alpha_s (Q_0))
=\exp\left[\int^{Q}_{Q_0}d\ln \mu\ \gamma_{\omega}(\alpha_s(\mu))\right].
\label{Eintermsofgamma}
\eeq
From the MLLA, the double and single logarithmic contribution to $\gamma_{\omega}(\alpha_s)$
reads
\beq
\gamma_{\omega}(\alpha_s)=\frac{1}{2}\left(-\omega + \sqrt{\omega^2+4\gamma_0^2}\right)
+\frac{\alpha_s}{2\pi}\left[b\frac{\gamma_0^2}{\omega^2+4\gamma_0^2}
-\frac{a}{2}\left(1+\frac{\omega}{\sqrt{\omega^2+4\gamma_0^2}}\right)\right]
+O\left(\left(\frac{\alpha_s}{\omega}\right)^3\right),
\label{MLLAformofanomdim}
\eeq
where $\alpha_s(Q)$ is calculated at one loop order and depends on the number of flavours 
$N_f$, $\gamma_0^2=4N_c \alpha_s/(2\pi)$ for $N_c=3$ colours,
$a=11 N_c/3+2N_f/(3N_c^2)$ and $b=11N_c/3-2N_f/3$. As is usual in applications of the
MLLA, we choose $N_f=3$. 
Eq. (\ref{MLLAformofanomdim}) is an expansion of $\gamma_{\omega}$ 
in $\alpha_s/\omega$ keeping $\alpha_s/\omega^2$ fixed. The first line is of order 
$\alpha_s/\omega$, and is obtained from the Double Logarithm Approximation (DLA), while the
second is the MLLA correction of $O\left(\left(\alpha_s/\omega\right)^2\right)$. 
The $O\left(\left(\alpha_s/\omega\right)^3\right)$
error is the unknown next-to-MLLA correction.
Since Eq.\ (\ref{MLLAformofanomdim}) reduces to a finite series in $\sqrt{\alpha_s}$ at $\omega=0$, 
the MLLA should also be a good approximation in the region $|\omega|,\sqrt{\alpha_s}\ll 1$ if the 
next-to-MLLA corrections in this region are of higher order in $\sqrt{\alpha_s}$.

One is interested in determining what improvements the MLLA makes to 
the cross section in $x$ space. This quantity can be obtained from
the inverse Mellin transform, given by
\beq
xD(x,Q)=\frac{1}{2\pi i}\int_C d\omega \exp[\omega \xi]D_{\omega}(Q),
\label{invLT}
\eeq
where the contour $C$ may be taken to be a straight line from
$\omega=\omega_0-i\infty$ to $\omega=\omega_0+i\infty$, where $\omega_0$ is real and to the right of all
singularities in $D_{\omega}(Q)$. Actually, the small $\xi$ dependence of $D(x,Q)$ is largely 
determined by $E_{\omega}$ at large $|\omega|$, which 
is described by the fixed order result. Coincidentally, 
the MLLA also leads to a good description of the small $\xi$ region, 
since $\gamma_{\omega}$ in Eq.\ (\ref{MLLAformofanomdim}) 
becomes negative at large $|\omega|$ like the fixed order result. On the other hand,
as $\xi\rightarrow \infty$ the contribution from the $|\omega|<O(\sqrt{\alpha_s})$ region
becomes increasingly relevant, and may eventually become dominant 
since the evolution of the cross section calculated in the fixed order approach 
falls off rapidly at large $|\omega|$ due to the $-\ln |\omega|$ behaviour of the anomalous dimensions 
to all orders. However, $\gamma_{\omega}$ in Eq.\ (\ref{MLLAformofanomdim}) only
approaches a constant, $-a\alpha_s/(2\pi)$, at large $|\omega|$,
so it cannot be expected that this approach to the MLLA gives a good description
at large $\xi$. There is no guarantee
that there exists some suitable choice for the large $|\omega|$ behaviour of
$D_{\omega}(Q_0)$ which can remedy this problem with the evolution.

Therefore we require an evolution which approximates 
the MLLA well in the $|\omega|<O(\sqrt{\alpha_s})$ region
and which falls off sufficiently fast at large $|\omega|$ such
that the small $|\omega|$ region gives the dominant contribution to the
cross section at large $\xi$. For this purpose we will study the 
MLLA in terms of the moments $K_n$ of the gluon FF, where
\beq
K_n(Q)=\left(-\frac{d}{d\omega}\right)^n\ln D_{\omega}(Q)\bigg{|}_{\omega=0},
\label{kappainM}
\eeq
since the first few moments ($n$ finite)
depend very little on the behaviour of the evolution at large $|\omega|$.
The $K$ moments completely determine $D_{\omega}(Q)$, since
Eq.\ (\ref{kappainM}) may be inverted using Taylor's Theorem to give, formally,
\beq
\ln D_{\omega}=\sum_{n=0}^{\infty} \frac{K_n(-\omega)^n}{n!}.
\label{expandlnMinnu}
\eeq
Note that the $K$ moments may be expressed in terms of the 
normalized $\xi$ moments, $\langle \xi^n \rangle$, which from Eq.\ (\ref{melofD}) 
can be calculated using
\beq
\langle \xi^n \rangle=\frac{1}{D_0}\left(-\frac{d }{d\omega}\right)^n D_{\omega}\bigg{|}_{\omega=0}.
\label{xininM}
\eeq
From Eqs.\ (\ref{defofanomdim}) and (\ref{kappainM}), the moments of the gluon FF evolve as
\beq
K_n(Q)=K_n(Q_0)+\Delta K_n(\alpha_s(Q),\alpha_s(Q_0)),
\label{delkapinindefint}
\eeq
where $\Delta K_n$ is defined to be the $n$th $K$ moment 
of the evolution $E_{\omega}$, and so the $K$ moments have the additional advantages that
their MLLA evolution is independent of $D_{\omega}(Q_0)$ and that
they each evolve independently of one another. This definition of $\Delta K_n$ together with
Eq.\ (\ref{Eintermsofgamma}) implies
\beq
\Delta K_n(\alpha_s(Q),\alpha_s  (Q_0))
=\int^Q_{Q_0} 
d\ln \mu \left(-\frac{d}{d\omega}\right)^n \gamma_{\omega}(\alpha_s(\mu))\bigg{|}_{\omega=0}.
\label{defofdelkap}
\eeq
Explicitly, Eqs. (\ref{MLLAformofanomdim}) and (\ref{defofdelkap}) for $n\geq 1$ give
\beq
\Delta K_n(\alpha_s(Q),\alpha_s(Q_0))
= \alpha_s^{-\frac{n+1}{2}}(Q)
\left(C_n^{(0)}+C_n^{(1)}\alpha_s^{\frac{1}{2}}(Q)+O\left(\alpha_s\right)\right)
-\big{\{}\alpha_s(Q) \leftrightarrow \alpha_s(Q_0)\big{\}},
\label{delkapatsmallalphas}
\eeq
where the $O\left(\alpha_s\right)$ error refers to unknown next-to-MLLA corrections, while
the $C_n^{(0,1)}$ are completely determined and 
are presented in \cite{Fong:1989qy} for the first few values of $n$. 
In fact, for $n\geq 3$ and odd, $C_n^{(0)}=0$. Eq.\ (\ref{delkapatsmallalphas})
also applies for $n=0$, but $\Delta K_0$ also contains a term proportional to $\ln \alpha_s$.

For small $|\omega|$, $E_{\omega}(\alpha_s(Q),\alpha_s(Q_0))$ may be approximated by
\beq
\ln E_{\omega}(\alpha_s(Q),\alpha_s(Q_0))=\sum_{n=0}^M 
\frac{\Delta K_n(\alpha_s(Q),\alpha_s(Q_0)) (-\omega)^n}{n!}
\label{evolwithhighmomsupp2}
\eeq
with $M$ finite. This corresponds to evolving the moments
$K_n$ for $n\leq M$ exactly as in the full unexpanded case, but fixing the remaining $K$ 
moments to be equal to $K_n(Q_0)$. Since the coefficient
$C_n^{(0)}$ in Eq.\ (\ref{delkapatsmallalphas}) for $n$ even
is positive when $n/2$ is odd and negative when $n/2$ is even, then
in the region where the imaginary part of $\omega$ is large,
if $M\geq 2$ and even, $Q$ and $Q_0$ are large and $Q_0< Q$, we obtain the fast decrease
\beq
E_{\omega} \rightarrow \exp\left[-\frac{\big{|}
\Delta K_M 
\omega^M\big{|}}{M!}\right].
\label{limbehavofEtrunc}
\eeq
If $M$ is odd, the $M$th term in Eq.\ (\ref{evolwithhighmomsupp2}) just produces an oscillation,
in which case the replacement $M\rightarrow M-1$ in Eq.\ (\ref{limbehavofEtrunc}) must be made.

It remains to be found whether Eq.\ (\ref{evolwithhighmomsupp2})
agrees well with the approach of Eqs.\ (\ref{defofanomdim}) to (\ref{MLLAformofanomdim})
in the whole region $|\omega|<O(\sqrt{\alpha_s})$.
In the extreme case $|\omega|=O(\sqrt{\alpha_s})$, Eq.\ (\ref{delkapatsmallalphas}) implies
\beq
\omega^n \Delta K_n=O\left(\alpha_s^{-\frac{1}{2}}\right),
\eeq
so that all terms in the series in Eq.\ (\ref{evolwithhighmomsupp2}) become of similar
magnitude. Such a series may still converge, or oscillate with an average value equal to
the unexpanded result. In any case, we see that the accuracy of Eq.\ (\ref{evolwithhighmomsupp2})
to reproduce the MLLA contribution to the cross section cannot be reliably determined
in Mellin space. Therefore we will try to determine what the suppression of higher moments
means in $x$ space. For this purpose, it will be
convenient to work with the moments $N$, $\overline{\xi}$, $\sigma^2$ and
$\kappa_n$ for $n=3,...,\infty$, defined by
\beq
\begin{split}
N=D_0,
\ \ \ \ \ 
\overline{\xi}=&\langle \xi \rangle,
\ \ \ \ \ 
\sigma^2=\langle \left(\xi -\overline{\xi}\right)^2\rangle, 
\ \ \ \ \ 
\kappa_3=\frac{\langle \left(\xi-\overline{\xi}\right)^3\rangle}{\sigma^3}, \\
\ \ \ \ \ 
\kappa_4=\frac{\langle \left(\xi-\overline{\xi}\right)^4\rangle}{\sigma^4}&-3,
\ \ \ \ \ 
\kappa_n=\frac{\langle \left(\xi-\overline{\xi}\right)^n\rangle}{\sigma^n}
\ \ \ \ \ (n\geq 5).
\end{split}
\label{defofusualmoms}
\eeq
Note that $\kappa_3$ is often written as $s$ and $\kappa_4$ as $k$.
From Eqs. (\ref{kappainM}) and (\ref{xininM}), these moments are related to the $K$ moments by
\beq
K_0=\ln N ,\ \ \ \ \ \ K_1=\overline{\xi},\ \ \ \ \ \ K_2=\sigma^2,\ \ \ \ \ \ 
K_n=\sigma^n \kappa_n\ \ \ \ \ (n\geq 3).
\label{kappatok}
\eeq
To obtain an expression for some function $D(x)$ in terms of the $\kappa_n$, 
we first make the replacement
$y=i\omega \sigma$ in Eq.\ (\ref{invLT}), which yields
\beq
xD(x)=\frac{N}{\sigma\sqrt{2\pi }}\exp\left[-\frac{\delta^2}{2}\right]
R(\delta,\{\kappa_n\}),
\label{FuptoksinvLT}
\eeq
where $\delta=(\xi -\overline{\xi})/\sigma$ and the real quantity $R$ is given by
\beq
R(\delta,\{\kappa_n\})
=\frac{e^{\delta^2/2}}{\sqrt{2\pi}}
\int_{-\infty}^{\infty}dy\exp\left[\sum_{n=3}^{\infty}\kappa_n\frac{(-iy)^n}{n!}\right]
\exp\left[iy\delta-\frac{y^2}{2}\right].
\label{defofR}
\eeq
Note that $R$ is equal to unity when all the $\kappa_n$ vanish.
We therefore see that a function depends on its moments $K_n$ for $n \geq 3$ only through the ratios
$\kappa_n=K_n/\sigma^n$. 
To calculate Eq.\ (\ref{FuptoksinvLT}) to a given accuracy when the $\kappa_n$ are small, we
expand the $\kappa_n$-dependent exponential in Eq.\ (\ref{defofR})
in powers of the $\kappa_n$ up to the required accuracy and perform the integral for each term. $R$
is then rewritten as an exponential of the form
\beq
R=\exp\left[\sum_{i=0}^\infty A_i \delta^i\right],
\label{generaldg}
\eeq
where each $A_i$ vanishes when all the $\kappa_n$ vanish. 
Now suppose the $\kappa_n$ are sufficiently small such that it is valid to expand
the $A_i$ in the $\kappa_n$ for $n\leq M$ up to some finite order, and 
neglect the moments $\kappa_n$ for $n>M$. In this case $\ln R$ as a series in
$\delta$ will terminate at some finite order, hence the 
argument of the exponential in Eq.\ (\ref{generaldg}) is 
not an expansion in $\delta$, since the approximation is valid even if $\delta$ is of $O(1)$.
For example, including the complete contribution from all terms of $O(s)$, $O(k)$,
$O(s^2)$, $O(k^2)$, $O(sk)$, $O(\kappa_5)$ and $O(\kappa_6)$ in $\ln R$ gives
\beq
\begin{split}
&xD(x)=\frac{N}{\sigma\sqrt{2\pi}} \exp\Bigg[\frac{1}{8}k-\frac{1}{2}s\delta
-\frac{1}{4}\left( 2+k \right)\delta^2+\frac{1}{6}s\delta^3+\frac{1}{24}k\delta^4\\
&-\frac{5}{24}s^2+\frac{1}{12} k^2-\frac{1}{48} \kappa_6 
+ \left( \frac{1}{8} \kappa_5-\frac{2}{3}sk\right) \delta
+ \left( \frac{1}{2} s^2-\frac{1}{3} k^2+\frac{1}{16} \kappa_6 \right) \delta^2\\
&+\left( -\frac{1}{12} \kappa_5+\frac{7}{12} sk \right) \delta^3
+ \left( -\frac{1}{8} s^2+\frac{7}{48} k^2-\frac{1}{48} \kappa_6\right) \delta^4\\
&+\left(\frac{1}{120} \kappa_5-\frac{1}{12} sk\right) \delta^5
+\left( -\frac{1}{72} k^2+\frac{1}{720} \kappa_6\right) \delta^6\Bigg].
\end{split}
\label{dgforFuptodel6}
\eeq

We now return to hadron production cross sections in the MLLA. In $x$ space, Eq.\ (\ref{defofanomdim})
becomes
\beq
xD(x,Q)=\int_x^1 \frac{dz}{z}
\frac{x}{z} E\left(\frac{x}{z},\alpha_s(Q),\alpha_s(Q_0)\right) z D(z,Q_0),
\eeq
where $E(z,\alpha_s(Q),\alpha_s(Q_0))$ is the inverse 
Mellin transform of $E_{\omega}(\alpha_s(Q),\alpha_s(Q_0))$.
From Eq.\ (\ref{kappatok}), the $\kappa$ moments of $zE(z,\alpha_s(Q),\alpha_s(Q_0))$,
$\kappa^E_n(\alpha_s(Q),\alpha_s(Q_0))$, obey (omitting arguments for brevity)
\beq
\kappa^E_n=\frac{\Delta K_n}{(\Delta K_2)^{\frac{n}{2}}}.
\label{delkapindelknodelsn}
\eeq
We may use Eq.\ (\ref{delkapatsmallalphas}) to expand $\kappa^E_n(\alpha_s(Q),\alpha_s(Q_0))$ 
as a series in $\alpha_s(Q)$ keeping $\alpha_s(Q_0)$ fixed, giving
\beq
\kappa^E_n(\alpha_s(Q),\alpha_s(Q_0)) \propto \alpha_s^{\frac{n-2}{4}}(Q) \left[1
+O\left(\alpha_s^{\frac{1}{2}}(Q)\right)\right].
\label{pertserfortilk}
\eeq
Therefore we may treat the $\kappa^E_n$ as small, in which case $E$ takes the form 
of Eq.\ (\ref{FuptoksinvLT}),
\beq
zE(z,\alpha_s(Q),\alpha_s(Q_0))
=\frac{N^E}{\sigma^E \sqrt{2\pi }}
\exp\left[-\frac{\delta^{E 2}(z)}{2}\right]
R^E(\delta^E(z),\{\kappa^E_n\}),
\eeq
where $\delta^E(z)=(\ln(1/z) -\overline{\xi}^E)/\sigma^E$.
Eq.\ (\ref{pertserfortilk}) and
the discussion after Eq.\ (\ref{generaldg}) show
that when $\ln R^E$ is expanded in $\alpha_s(Q)$ while
$\delta^E(z)$ is fixed, the higher moments
serve only to contribute spurious higher order terms to this
series. Such terms give a large theoretical error, since
they contribute unstable ``noise'', as can be seen by performing the evolution
on a smooth function. Discarding these terms gives an evolution in which
the higher moments of the gluon FF are fixed with respect to $Q$, in other words Eq.\
(\ref{evolwithhighmomsupp2}). 

Similarly, the $\kappa$ moments of $D(x,Q)$ obey
\beq
\kappa_n(Q)=\frac{K_n(Q_0)+\Delta K_n(\alpha_s(Q),\alpha_s(Q_0))}
{\left[K_2(Q_0)+\Delta K_2(\alpha_s(Q),\alpha_s(Q_0))\right]^{\frac{n}{2}}}.
\label{kapninKsiganddels}
\eeq
At sufficiently large $Q$, $\Delta K_n(\alpha_s(Q),\alpha_s(Q_0)) \gg K_n(Q_0)$,
so that, from Eqs.\ (\ref{delkapindelknodelsn}) and (\ref{kapninKsiganddels}),
$\kappa_n(Q)$ may be approximated by $\kappa^E_n(\alpha_s(Q),\alpha_s(Q_0))$,
and hence may be treated as small.
Thus if $Q_0$ is sufficiently large, $xD(x,Q_0)$ may be parameterized as a distorted Gaussian
around the average value of $\xi$.

\section{Fitting to data}
\label{FtD}

\begin{figure*}[ht!]
\centering
\setlength{\epsfxsize}{18cm}
\begin{minipage}[ht]{\epsfxsize}
\centerline{\mbox{\epsffile{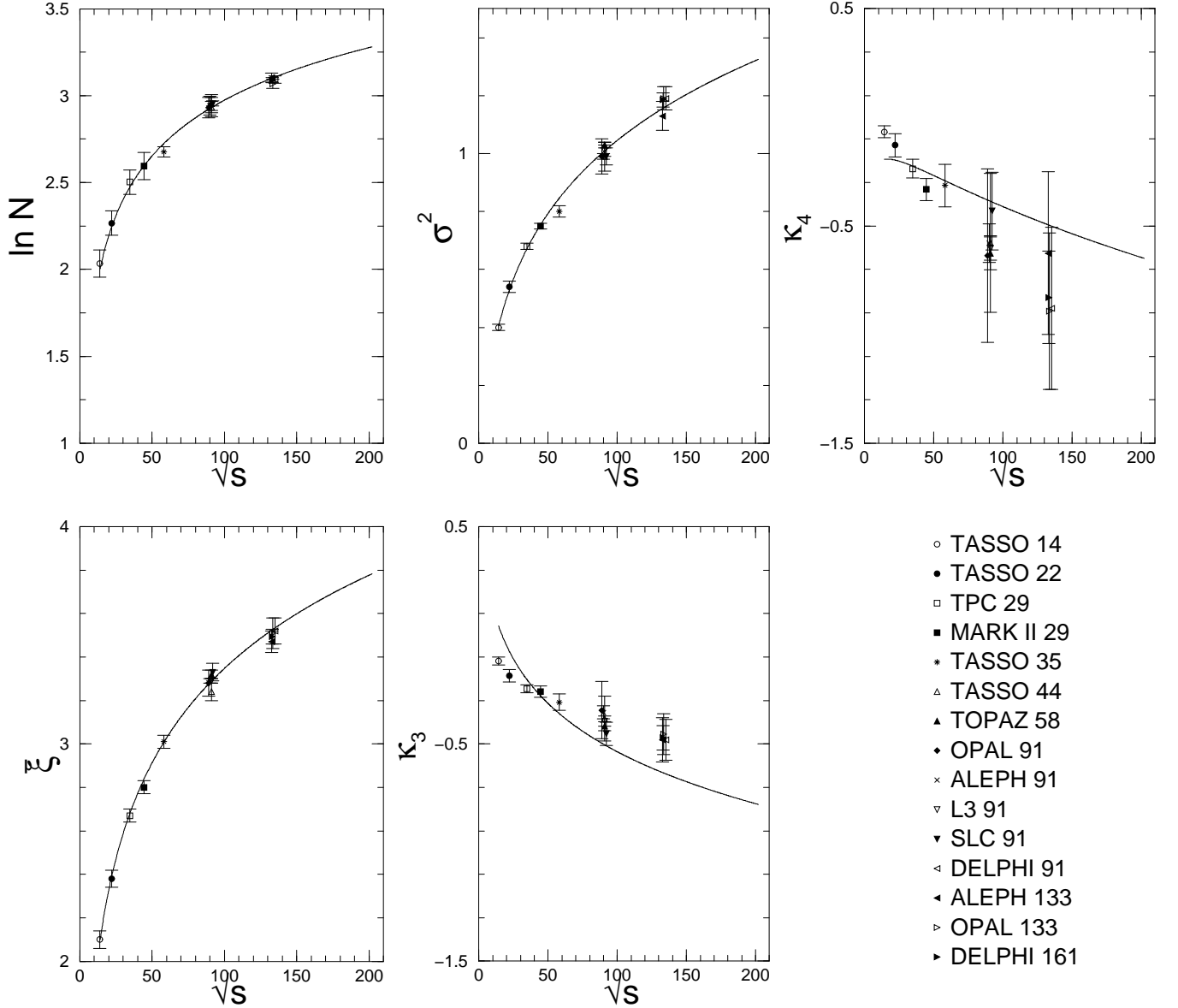}}}
\end{minipage}
\caption{Simultaneous fit of the initial moments and $\Lambda_{\rm QCD}$ to all of the 
experimental data moments calculated in \cite{Lupia:1997hj}, by evolving the moments
in the MLLA.
\label{fig7}}
\end{figure*}

Since the evolution of the first few $K$ moments is independent of the approach used 
(e.g. the approach of Eqs. (\ref{defofanomdim}) to (\ref{MLLAformofanomdim}) or the
approach of Eq.\ (\ref{evolwithhighmomsupp2})) and the parameterization of the non-perturbative
input, the ability of the MLLA to describe the data in principle can be 
determined by comparison to the moments of the data, assuming that 
fixed order corrections can be neglected.
We perform a single fit of $\Lambda_{\rm QCD}$ and the initial $K_n$ with $Q_0=14/2$ GeV to the 
$K_n$ of the experimental data at various $\sqrt{s}$ by setting
$Q=\sqrt{s}/2$ and evolving in the MLLA as in Eq.\ (\ref{defofdelkap}). We use the moments
calculated in \cite{Lupia:1997hj}, which are presented in the form
$N$, $\overline{\xi}$, $\sigma^2$, $s$ and $k$, from which we extract the first
5 $K$ moments, and their errors are obtained
by differentiating and adding in quadrature. The results are shown in Table \ref{tab7} and Fig.\ \ref{fig7}.
The error on $\Lambda_{\rm QCD}$ in Table \ref{tab7} and in
all other tables in this paper is obtained by inverting the correlation
matrix, which is identified with the matrix of second derivatives.
In Fig.\ \ref{fig7}, we see that
the first 3 moments are fitted very well, however there is a marginal disagreement of $K_3$
and $K_4$ with the data which may result from either an inability 
of the MLLA to describe these higher $K$ moments
or from the inaccuracy involved in obtaining these moments from the
experimental data, in which case the true experimental error would be 
larger than that shown. We also perform a fit directly
to the basis of moments used in \cite{Lupia:1997hj}, and found no significant change in the 
values of the initial parameters, and the theoretical curves for $s$ and $k$ deviated seriously
from the data below $\sqrt{s}\approx 25$ GeV.
\begin{table}
\caption{Simultaneous fit of $\Lambda_{\rm QCD}$ and the moments
$\ln N$, $\overline{\xi}$, $\sigma^2$, $K_3$ and
$K_4$ at $Q_0=14/2$ GeV to the same moments of 
the data from \cite{Lupia:1997hj}. $\chi^2_{\rm DF}=3.7$.
\label{tab7}}
\begin{tabular}{llllll}
\hline\noalign{\smallskip}
$\ln N$\ \ \ \ \ \  & $\overline{\xi}$\ \ \ \ \ \ \ \ & $\sigma^2$\ \ \ \ \ \  & $K_3$\ \ \ \ \ \ 
& $K_4$\ \ \ \ \ \ & 
$\Lambda_{\rm QCD}$ (MeV)\\
\noalign{\smallskip}\hline\noalign{\smallskip}
2.00 & 2.09 & 0.40 & 0.05 & -0.19 & 411\ $\pm$ 36 \\
\noalign{\smallskip}\hline
\end{tabular}
\end{table}

These results strongly suggest that there exists some approach to applying the MLLA
and some parameterization of the non-perturbative components that gives a good description
of the data over a large range in $\xi$. In the following three subsections we perform
fits directly to the data points at different $\xi$ using various approaches. In Subsection \ref{EDG},
we evolve the moments, place them in a distorted Gaussian and compare with the
data at different $\xi$. In Subsections \ref{MSP} and \ref{DGwMSE},
we evolve in Mellin space
using Eq.\ (\ref{evolwithhighmomsupp2}), with two different parameterizations of the
non-perturbative input: In Subsection \ref{MSP} we parameterize the initial distribution 
$xD(x,Q_0)$ such that its higher moments are exactly zero but the remaining moments are left as free
parameters, while in Subsection \ref{DGwMSE} we parameterize the initial distribution
as a distorted Gaussian (the first two lines of Eq.\ (\ref{dgforFuptodel6})) so that
the higher moments are small. In order to avoid the small $\xi$ region, 
where fixed order effects are important and where the data have a high accuracy,
we follow \cite{Abbiendi:2002mj} and use only those data for which
\beq
\xi >0.75+0.33\ln(\sqrt{s}),
\label{OPALlowercutonxi}
\eeq
but impose no upper limit in $\xi$ on the data. 

\subsection{Evolution of moments as distorted Gaussian parameters}
\label{EDG}

\begin{figure}[ht!]
\centering
\setlength{\epsfxsize}{8.5cm}
\begin{minipage}[ht]{\epsfxsize}
\centerline{\mbox{\epsffile{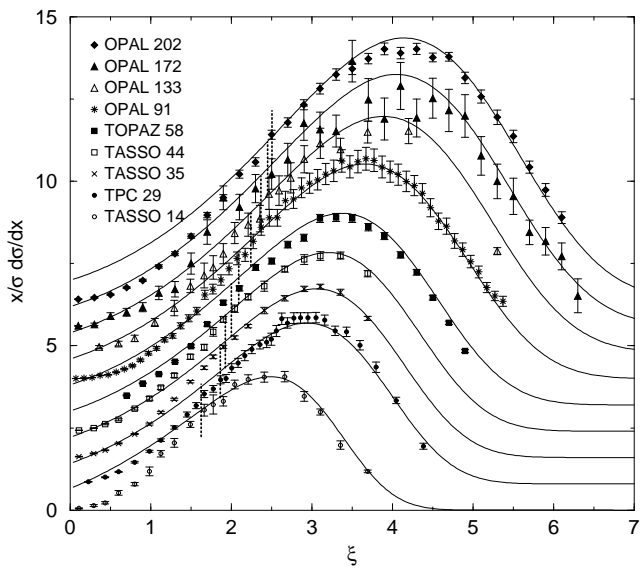}}}
\end{minipage}
\caption{Global fit by using a distorted Gaussian in which the moments are evolved in the MLLA.
The lower limits of the data used, given by Eq.\ (\ref{OPALlowercutonxi}), are indicated by vertical 
dotted lines. Each curve is shifted up by 0.8 for clarity.
\label{fig10}}
\end{figure}

While the data tend to follow the shape of a distorted Gaussian well, the 
distorted Gaussian evolved with the approach of Eqs. (\ref{defofanomdim}) to (\ref{MLLAformofanomdim})
does not --- in \cite{Albino:2004xa}, the theoretical curves exhibited two bumps. 
Therefore we constrain the evolved $xD(x,Q)$ to follow a distorted Gaussian, 
\beq
xD(x,Q)=\frac{N'}{\sigma^{\prime}\sqrt{2\pi}}
\exp\Bigg[\frac{1}{8}k'-\frac{1}{2}s'\delta^{\prime}
-\frac{1}{4}\left( 2+k' \right)\delta^{\prime 2}
+\frac{1}{6}s'\delta^{\prime 3}+\frac{1}{24}k'\delta^{\prime 4}\Bigg],
\eeq
with $\delta^{\prime}=(\xi-\overline{\xi}^{\prime})/\sigma^{\prime}$,
where the parameters $N'$, $\overline{\xi}^{\prime}$, $\sigma^{\prime}$, $k'$ and $s'$
depend on $Q$. Since these parameters are approximately equal to the corresponding 
unprimed quantities defined in Eq.\ (\ref{defofusualmoms}), we choose their $Q$ dependences to be the same,
i.e.\ that obtained from Eqs.\ (\ref{defofdelkap}) and (\ref{kappatok}).
By comparing $xD(x,Q)$ calculated in this way
with the data at different $\xi$ and $\sqrt{s}$, we fit the initial 
$K_n^{\prime}(Q_0)=K_n^{\prime}$ for $n\leq 5$, as well as $\Lambda_{\rm QCD}$, with the choice 
$Q_0=14/2$ GeV. A similar approach was applied in \cite{Abbiendi:2002mj}, however
the Limiting Spectrum formulae for the moments were used, and
$\Lambda_{\rm QCD}$, the peak position of the data and the normalisation
of the data for each $\sqrt{s}$ were fitted. 
We use TASSO data at $\sqrt{s}=$14, 22, 35 and 44 GeV \cite{Braunschweig:1990yd},
TPC \cite{Aihara:1988su} and MARK II \cite{Petersen:1987bq} data at 29 GeV,
TOPAZ data at 58 GeV \cite{Itoh:1994kb},
ALEPH \cite{Barate:1996fi}, DELPHI \cite{Abreu:1996na}, L3 \cite{Adeva:1991it},
OPAL \cite{Akrawy:1990ha} and SLD \cite{Abe:1998zs} data at 91 GeV,
ALEPH \cite{Buskulic:1996tt} and OPAL \cite{Alexander:1996kh} data at 133 GeV,
DELPHI data at 161 GeV \cite{Ackerstaff:1997kk},
OPAL data at 172, 183 and 189 GeV \cite{Abbiendi:1999sx} 
and OPAL data at 202 GeV \cite{Abbiendi:2002mj}.
The results are shown in Table \ref{tab10}, and some of the data
with the corresponding fitted theoretical curves
are shown in Fig.\ \ref{fig10}. 
(In all plots in this paper, each curve is shifted up from the curve below by 0.8 for clarity.)
The fit is excellent around the peak and above, but poor below
the peak region, where fixed order effects are important. 
In particular, the curves show that the MLLA alone predicts the 
evolution of the normalization very well at large $\xi$,
in contrast to other analyses. 
\begin{table}
\caption{Simultaneous fit of $\Lambda_{\rm QCD}$ and the moments
$\ln N'$, $\overline{\xi}'$, $\sigma^{\prime 2}$, $K_3^{\prime}$ and
$K_4^{\prime}$ at $Q_0=14/2$ GeV to the data in $x$ space by evolving 
them in the MLLA and placing them in a distorted Gaussian. 
$\chi^2_{\rm DF}=2.4$.
\label{tab10}}
\begin{tabular}{llllll}
\hline\noalign{\smallskip}
$\ln N'$\ \ \  \ \ \  & $\overline{\xi}'$\ \ \ \ \ \ \ \ & $\sigma^{\prime 2}$\ \ \  \ \ \  & 
$K_3^{\prime}$ \ \ \  \ \ \ 
& $K_4^{\prime}$ \ \ \  \ \ \ & $\Lambda_{\rm QCD}$ (MeV)\\
\noalign{\smallskip}\hline\noalign{\smallskip}
2.41 & 2.07 & 0.91 & -0.58 & -0.66 & 59\ $\pm$ 4 \\
\noalign{\smallskip}\hline
\end{tabular}
\end{table}
This procedure has a number of features that differ from that
used in \cite{Albino:2004xa}. The evolution
used there coincidentally shares similar properties at large $|\omega|$ with
the fixed order result, and hence a good fit was obtained with the data below the peak while
a large disagreement was found above the peak. However, no properties of the evolution used for the fit of 
Table \ref{tab10} are shared
with the fixed order result below the peak, and only the first 5 moments are evolved while the 
remaining moments are held fixed (within the accuracy of the distorted Gaussian approximation). 
To ensure that the deviation below the peak
was not due to the lack of data being fitted there, a fit using all data for which
$\xi<\ln (\sqrt{s}/20.5\ {\rm GeV})$ was performed, giving a bad fit
everywhere. In particular since the parameters try to fit to
the accurate data below the peak, a large initial $|K_4|$ was obtained.

\subsection{Mellin space parameterization and evolution}
\label{MSP}

We now constrain the initial distribution such that $K_n=0$ exactly 
for $n\geq 7$. From Eq.\ (\ref{expandlnMinnu}), this means that the
parameterization of the initial distribution in Mellin space must take the form
\beq
\ln D_{\omega}(Q_0)=\sum_{n=0}^6 \frac{K_p(Q_0)(-\omega)^n}{n!}.
\eeq
We perform fits to TASSO data at 14 GeV and OPAL data at 91 and 202 GeV, 
cut according to Eq.\ (\ref{OPALlowercutonxi}), and
evolve according to Eq.\ (\ref{evolwithhighmomsupp2}) with $Q_0=14/2$ GeV. Taking $M=$ 2, 4 and 6,
we obtain the results shown in Tables \ref{tabfb24}, \ref{tabfb44} and \ref{tabfb64}
respectively. Note that using $M=4$ gives the lowest $\chi^2_{\rm DF}$.
For the case $M=6$, we get a large result for $\kappa_6$, since 
although $\Delta K_6(\alpha_s(Q),\alpha_s(Q_0))$ becomes positive for $Q\rightarrow \infty$,
for the data used $\Delta K_6$ is negative. This is due to the MLLA term being
larger than the DLA term, so it may be the case that corrections beyond the MLLA are required
for this quantity, or some other approach. At any rate, from the end of Section \ref{MitMLLA}, 
$\Delta K_6$ contributes spurious higher order terms and should be neglected,
and this is confirmed in Table \ref{tabfb64}.
\begin{table*}
\caption{Simultaneous fit of $N$, $\overline{\xi}$, $\sigma^2$, $s$,
$k$, $\kappa_5$, $\kappa_6$ and $\Lambda_{\rm QCD}$ to
TASSO data at 14 GeV and OPAL data at 91 and 202 GeV, 
with only the first 
3 moments evolved according to the MLLA.
$\chi^2_{\rm DF}=1.14$.
\label{tabfb24}}
\begin{tabular}{llllllll}
\hline\noalign{\smallskip}
$N$\ \ \ \ \ \  & $\overline{\xi}$ \ \ \ \ \ \ \ & $\sigma^2$\ \ \ \ \ \ & $s$\ \ \ \ \ \ \ 
& $k$\ \ \ \ \ \ \ & $\kappa_5$\ \ \ \ \ \ \ & $\kappa_6$\ \ \ \ \ \ \ & $\Lambda_{\rm QCD}$ (MeV)\\
\noalign{\smallskip}\hline\noalign{\smallskip}
10.90 & 2.40 & 1.05 & 0.05 & 0.54 & 0.09 & 0.19 & 66\ $\pm$ 10 \\
\noalign{\smallskip}\hline
\end{tabular}
\end{table*}
\begin{table*}
\caption{As in Table \ref{tabfb24}, but with only the first 
5 moments evolved according to the MLLA. $\chi^2_{\rm DF}=0.60$.
\label{tabfb44}}
\begin{tabular}{llllllll}
\hline\noalign{\smallskip}
$N$\ \ \ \ \ \  & $\overline{\xi}$\ \ \ \ \ \ \ \ & $\sigma^2$\ \ \ \ \ \ \ & $s$\ \ \ \ \ \ \ 
& $k$\ \ \ \ \ \ \ & $\kappa_5$\ \ \ \ \ \ \ & $\kappa_6$\ \ \ \ \ \ & $\Lambda_{\rm QCD}$ (MeV)\\
\noalign{\smallskip}\hline\noalign{\smallskip}
10.58 & 2.15 & 0.79 & -0.79 & -0.40 & -0.73 & -2.44 & 81\ $\pm$ 11 \\
\noalign{\smallskip}\hline
\end{tabular}
\end{table*}
\begin{table*}
\caption{As in Table \ref{tabfb24}, but with only the first 
7 moments evolved according to the MLLA. $\chi^2_{\rm DF}=0.77$.
\label{tabfb64}}
\begin{tabular}{llllllll}
\hline\noalign{\smallskip}
$N$\ \ \ \ \ \ \ \ & $\overline{\xi}$\ \ \ \ \ \ \ & $\sigma^2$\ \ \ \ \ \ \ \ & $s$\ \ \ \ \ \ \ \ \ \ 
& $k$\ \ \ \ \ \ \ \ \ \ & $\kappa_5$\ \ \ \ \ \ \ \ \ \ 
& $\kappa_6$\ \ \ \ \ \ \ \ \ & $\Lambda_{\rm QCD}$ (MeV)\\
\noalign{\smallskip}\hline\noalign{\smallskip}
10.06 & 1.80 & 0.474 & -4.86 & 2.00 & -1.17 & 113.57 & 96\ $\pm$ 1 \\
\noalign{\smallskip}\hline
\end{tabular}
\end{table*}

\subsection{Distorted Gaussian with Mellin space evolution}
\label{DGwMSE}

\begin{figure}[ht!]
\centering
\setlength{\epsfxsize}{8.5cm}
\begin{minipage}[ht]{\epsfxsize}
\centerline{\mbox{\epsffile{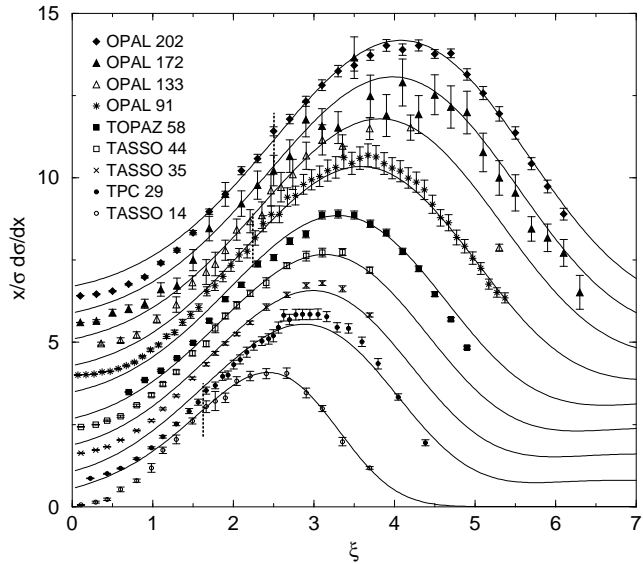}}}
\end{minipage}
\caption{Fit of the distorted Gaussian parameters and 
$\Lambda_{\rm QCD}$ to TASSO data at 14 GeV and OPAL data at 91 and 202 GeV, 
with only the first 
5 moments evolved according to the MLLA. 
\label{fig13}}
\end{figure}

In global analyses, the initial FF's are parameterized in $x$, and the parameters
are fitted to data by evolving the FF's in Mellin space in the fixed order approach. 
For this reason, in \cite{Albino:2004xa}
the initial gluon FF was parameterized as a distorted Gaussian and evolved
using the approach of Eqs. (\ref{defofanomdim}) to (\ref{MLLAformofanomdim}).
We repeat this approach here, but instead 
we will apply MLLA evolution in the form of Eq.\ (\ref{evolwithhighmomsupp2}).
First we repeat the fits of Subsection \ref{MSP}.
The results for the case $M=2$ and $M=4$ are shown in Tables \ref{tabfh65} and
\ref{tab13} respectively. Note again that using $M=4$ gives the best fit.
The resulting curves for the $M=4$ case are shown in Fig.\ \ref{fig13}.
The cases $M=6$ cannot be tested since, as we found in Subsection \ref{MSP},
$\Delta K_6$ is negative for the data used so that the integral for the 
inverse Mellin transform does not converge.
Fits in which data for all $\xi$ is used generally give a bad fit everywhere except at
values of $\xi$ beyond the peak, which is due in part to the high accuracy of the data at small $\xi$.
As anticipated from the fit of Table \ref{tab10}, 
excellent agreement is found around and above the peak region. 
\begin{table}
\caption{Simultaneous fit of $N'$, $\overline{\xi}'$, $\sigma^{\prime 2}$, $s'$,
$k'$ and $\Lambda_{\rm QCD}$ to
TASSO data at 14 GeV and OPAL data at 91 and 202 GeV, 
with only the first 
3 moments evolved according to the MLLA. $\chi^2_{\rm DF}=1.67$.
\label{tabfh65}}
\begin{tabular}{llllll}
\hline\noalign{\smallskip}
$N'$\ \ \ \ \ \ \ \ \  & $\overline{\xi}'$\ \ \ \ \ \ \ \ \ 
& $\sigma^{\prime 2}$\ \ \ \ \ \ \ \ \ 
& $s'$\ \ \ \ \ \ \ \ \ 
& $k'$\ \ \ \ \ \ \ \ \ & $\Lambda_{\rm QCD}$ (MeV)\\
\noalign{\smallskip}\hline\noalign{\smallskip}
10.51 & 2.22 & 1.17 & -0.71 & 1.18 & 86\ $\pm$ 1 \\
\noalign{\smallskip}\hline
\end{tabular}
\end{table}
\begin{table}
\caption{As in Table \ref{tabfh65}, but with only the first 
5 moments evolved according to the MLLA. $\chi^2_{\rm DF}=0.65$.
\label{tab13}}
\begin{tabular}{llllll}
\hline\noalign{\smallskip}
$N'$\ \ \ \ \ \ \ & $\overline{\xi}'$\ \ \ \ \ \ \ \ \ 
& $\sigma^{\prime 2}$\ \ \ \ \ \ \ \ \ 
& $s'$\ \ \ \ \ \ \ \ \ 
& $k'$\ \ \ \ \ \ \ \ \ & $\Lambda_{\rm QCD}$ (MeV)\\
\noalign{\smallskip}\hline\noalign{\smallskip}
10.75 & 2.12 & 0.92 & -0.65 & -0.22 & 77\ $\pm$ 10 \\
\noalign{\smallskip}\hline
\end{tabular}
\end{table}

\begin{figure}[ht!]
\centering
\setlength{\epsfxsize}{8.5cm}
\begin{minipage}[ht]{\epsfxsize}
\centerline{\mbox{\epsffile{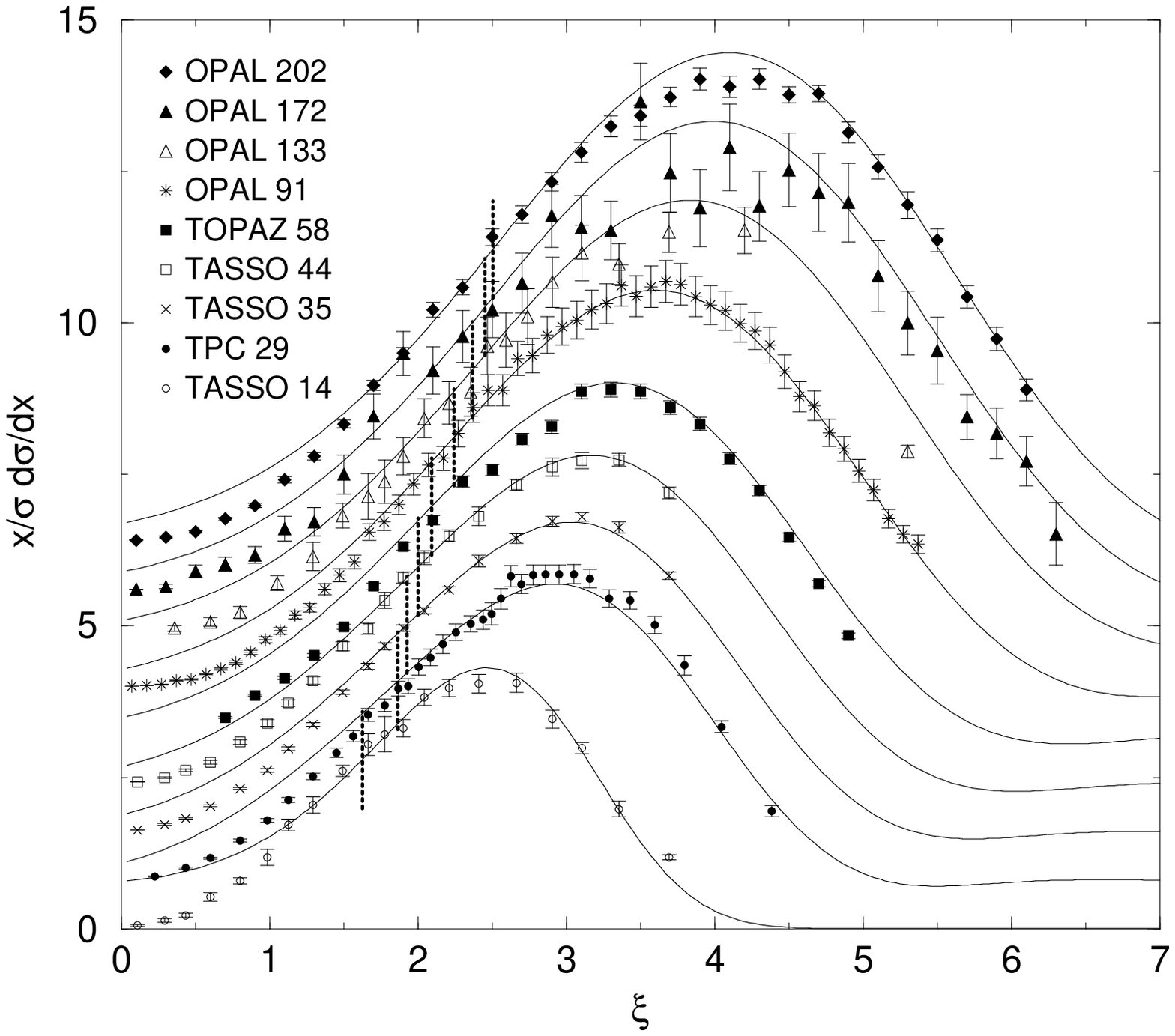}}}
\end{minipage}
\caption{Global fit with only the first 5 moments evolved according to the MLLA.
\label{figfg91}}
\end{figure}

We now perform a fit to all the data, using $M=4$ in the evolution.
The results are shown in Table
\ref{tabfg91} and Fig.\ \ref{figfg91}, and are the main results of this paper. 
\begin{table}
\caption{Global fit with only the first 5 moments evolved according to the MLLA. 
$\chi^2_{\rm DF}=2.7$.
\label{tabfg91}}
\begin{tabular}{llllll}
\hline\noalign{\smallskip}
$N'$\ \ \ \ \ \ \ \ \ & $\overline{\xi}'$\ \ \ \ \ \ \ \ \ & $\sigma^{\prime 2}$\ \ \ \ \ \ \ \ \ 
& $s'$\ \ \ \ \ \ \ \ \ 
& $k'$\ \ \ \ \ \ \ \ \ & $\Lambda_{\rm QCD}$ (MeV)\\
\noalign{\smallskip}\hline\noalign{\smallskip}
10.39 & 1.90 & 1.10 & -1.45 & -0.09 & 114\ $\pm$ 6 \\
\noalign{\smallskip}\hline
\end{tabular}
\end{table}

In all our approaches, reasonably consistent values of $\Lambda_{\rm QCD}$ were obtained of
around 100 MeV. The small $\xi$ region was generally not well described, however this
region is outside the scope of the MLLA and requires fixed order corrections. 
It is generally found that the best fit is obtained with $M=4$, although with other values
of $M$ we also obtained good fits around and above the peak.

\section{The LPHD and Limiting Spectrum limits}
\label{LPHDLS}

From Eq.\ (\ref{delkapatsmallalphas}), $\Delta K_n(\alpha_s(Q),\alpha_s(Q_0))\rightarrow \infty$
as $Q\rightarrow \infty$, so that for sufficiently large $Q$ the initial $K_n(Q_0)$ 
in Eq.\ (\ref{delkapinindefint}) may be neglected. However, such an approximation
should not be made for $n=0$, since the cross section is very sensitive to the initial $\ln N$.
Therefore, data at sufficiently large $Q$ should be reasonably well described with a
starting distribution of the form $xD(x,Q_0)=N\delta(1-x)$. Thus, from a
purely perturbative analysis, we see that the LPHD arises because the
perturbative components form the dominant contribution to the cross
section. In this sense, the LPHD follows from the MLLA rather than being an
additional assumption. 

For sufficiently large $Q$, any term of $O(\alpha_s^{-n}(Q_0))$ with $n$ positive
can be neglected relative to a term of $O(\alpha_s^{-n}(Q))$, so that from 
Eq.\ (\ref{delkapatsmallalphas}), we may replace $\Delta K_n(\alpha_s(Q),\alpha_s(Q_0))$
with $\Delta K_n(\alpha_s(Q),\infty)$. This can be artificially achieved by setting
$Q_0=\Lambda_{\rm QCD}$, since then 
$\alpha_s(Q_0)=\infty$ as a consequence of the perturbative approximation.
Thus the Limiting Spectrum also follows from the MLLA.
However, it is important to note that the $\Delta K_n(\alpha_s(Q),\infty)$ 
for $n$ sufficiently small would not be finite if
corrections of next-to-MLLA or higher were included. 

\begin{figure}[ht!]
\centering
\setlength{\epsfxsize}{8.5cm}
\begin{minipage}[ht]{\epsfxsize}
\centerline{\mbox{\epsffile{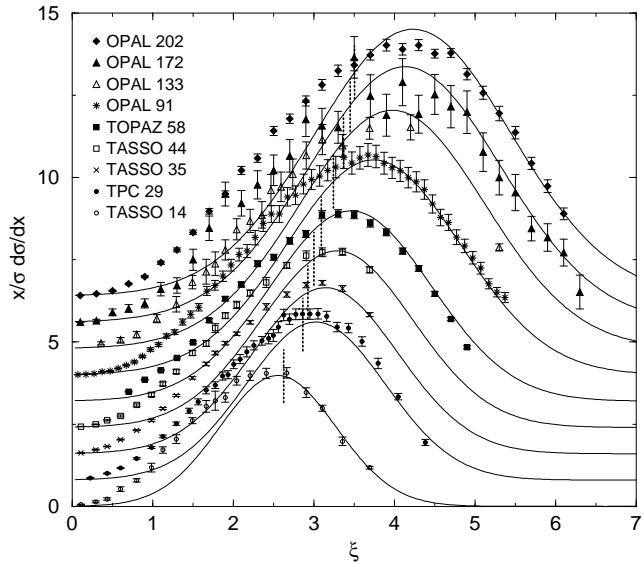}}}
\end{minipage}
\caption{Global fit to data with the LPHD and Limiting Spectrum, and with
only the first 3 moments evolved.
\label{figfw91}}
\end{figure}

To summarize, a simple analysis of the MLLA shows that, at sufficiently large $Q$, 
the assumptions of the LPHD and the Limiting Spectrum
will appear to be correct, since the cross section
may be calculated with Eq.\ (\ref{delkapinindefint}) approximated for $n\geq 1$ by
\beq
K_n(Q)\approx \Delta K_n(\alpha_s(Q),\infty).
\label{LPHDLSlimit}
\eeq
To properly test the physical assumptions which imply the
LPHD and the Limiting Spectrum requires a comparison with data in the region
$Q = O(\Lambda_{\rm QCD})$.
However, the only essential difference between our procedure and that of the LPHD
with the Limiting Spectrum is that we do not allow for any assumptions
on the size of the initial $\overline{\xi}$ and $\sigma^2$, we only assume
that the initial $\kappa_n$ are small.
We now study what the effects of these additional constraints are, by 
using Eq.\ (\ref{LPHDLSlimit}) for $n\geq 1$. From Eq.\ (\ref{delkapinindefint}),
the normalization $N(Q)$
obeys $\ln N(Q)=\ln \overline{N}+\Delta K_0(\alpha_s(Q),\infty)$, where
$\ln \overline{N}=\ln N(Q_0)-\Delta K_0(\alpha_s(Q_0),\infty)$ is independent of $Q_0$.
Thus our predictions will be exactly independent of $Q_0$, and the theory 
contains just two free parameters, $\overline{N}$ and $\Lambda_{\rm QCD}$, which we
fit to all available data. We find that the data above the cut is best described when the cut
is chosen as in Eq.\ (\ref{OPALlowercutonxi}) but with
$\Delta \xi= +1$ added to the right hand side. Fitting a different normalization
for each $\sqrt{s}$ of the data does not improve the fit significantly.
For $M=4$ in the evolution, 
we obtain $\overline{N}=7.68 \pm 0.02$, $\Lambda_{\rm QCD}=446 \pm 2$ and $\chi^2_{\rm DF}=9.3$,
while for $M=2$ we obtain $\overline{N}=6.64 \pm 0.02$, 
$\Lambda_{\rm QCD}=292 \pm 1$ GeV and $\chi^2_{\rm DF}=3.74$.
The resulting plots are shown in Fig.\ \ref{figfw91} for the $M=2$ case. (The $M=4$
fit gives similar curves.) With these results we are able to calculate
the moments at $Q=14/2$ GeV, and find reasonable agreement with our previous results for
$N$, $\overline{\xi}$ and $\sigma^2$ (e.g.\ that of Table \ref{tabfg91}). The evolution
of these quantities follows the data, as is to be expected since we are using the same
evolution as we used in the fit of, e.g., Table \ref{tabfg91}.
However, it is clear that the prediction at and below the peak is significantly dependent on
the initial values of $\xi$ and $\sigma$.
Comparing with other analyses which use the LPHD with the Limiting Spectrum, 
we see that the data above the peak are well described, with no modifications
to the normalization, while the data below the peak are not. 

\section{Conclusions}
\label{conclusions}

In \cite{Albino:2004xa}, it was shown that a naive application
of the MLLA without additional assumptions gives a good description of
data around the peak region, but not beyond. Since
the evolution used approached a constant at large $|\omega|$, while
the fixed order approach implies that the evolution falls to zero,
the contribution to the cross section at large $\xi$ from the cross section 
at small $|\omega|$ was underestimated. This qualitative feature of the
fixed order approach can be taken into account by using the Limiting
Spectrum, which decreases fast at large $|\omega|$. In this paper, we found that this 
feature can also be taken into acount by suppressing the evolution of the higher moments. This
does not affect the approximation since, by studying the MLLA in $x$ space, one
can see that it is formally justified to neglect their evolution in the perturbative
approximation. In addition, at sufficiently large $Q$, 
one finds from the MLLA that the spectrum acquires a distorted Gaussian
shape. Fixing the higher moments and using a distorted Gaussian for the 
initial distribution at $Q_0=14/2$ GeV resulted in a good description of 
all data for which $\sqrt{s}\geq 14$ GeV from just below the peak to the 
largest value of $\xi$, and one obtains $\Lambda_{\rm QCD}\approx 100$ GeV.
In obtaining these results, we used a non-perturbative input that was determined
empirically at a low scale, rather than from physical arguments such as those of the LPHD.

Furthermore, we showed that the cross section approaches that of the 
Limiting Spectrum at sufficiently large $Q$. We stress that this follows from the MLLA,
without additional hypotheses. In practice, imposing the Limiting Spectrum limit on the 
evolution with suppressed moments and imposing the LPHD on the initial distribution gives
a poor description up to the peak. However, a good description is obtained
beyond the peak without requiring any modification to the normalization, and 
the best fit, where $M=2$, resulted in $\Lambda_{\rm QCD}=292$ GeV, which is consistent
with other analyses.

Finally, the region of the data used in global fits may be extended to
lower values of $x$ (large $\xi$) by incorporating the MLLA into the fixed order 
calculations. This would allow for a better determination of the
FF's, particularly at small momentum fractions, as well as more constraints on $\Lambda_{\rm QCD}$.

\section*{Acknowledgments}

The authors would like to thank Wolfgang Ochs for helpful suggestions.
This work was supported in part by the Deutsche Forschungsgemeinschaft     
through Grant No.\ KN~365/1-2, by the Bundesministerium f\"ur Bildung und  
Forschung through Grant No.\ 05~HT4GUA/4, and by Sun Microsystems through  
Academic Equipment Grant No.\ EDUD-7832-000332-GER.




\end{document}